\pdfoutput=1
\documentclass[prd,twocolumn,preprintnumbers,superscriptaddress,nofootinbib,showpacs]{revtex4}
\usepackage[a4paper, hdivide={1.91cm,,1.165cm}, vdivide={1.83cm,,3.8cm}]{geometry}

\usepackage{amsmath}
\usepackage{graphicx}
\usepackage{dcolumn}
\usepackage[hyperfootnotes=false]{hyperref}
\usepackage{units}

\newcommand{\diag}{\mathrm{diag}\,}

\newcommand{\Br}{\text{Br}}
\newcommand{\tr}{\text{tr}}

\def \epsilon {\varepsilon} 

\def \vec#1{{\boldsymbol{#1}}}
\newcommand{\matrixx}[1]{\begin{pmatrix} #1 \end{pmatrix}} 
\newcommand{\dd}{\mathrm{d}}
\newcommand{\M}{\mathcal{M}}

\allowdisplaybreaks 

\begin{document}

\hypersetup{
    pdftitle={Scalar glueballs: Constraints from the decays into \texorpdfstring{$\eta$}{eta} or \texorpdfstring{$\eta'$}{eta-prime}},
    pdfauthor={Jean-Marie Frere, Julian Heeck}
}

\title{Scalar glueballs: Constraints from the decays into \texorpdfstring{$\eta$}{eta} or \texorpdfstring{$\eta'$}{eta-prime}}

\author{Jean-Marie \surname{Fr\`ere}}
\email{frere@ulb.ac.be}
\affiliation{Service de Physique Th\'eorique, Universit\'e Libre de Bruxelles, Boulevard du Triomphe, CP225, 1050 Brussels, Belgium}

\author{Julian \surname{Heeck}}
\email{julian.heeck@ulb.ac.be}
\affiliation{Service de Physique Th\'eorique, Universit\'e Libre de Bruxelles, Boulevard du Triomphe, CP225, 1050 Brussels, Belgium}

\preprint{ULB-TH/15-09}



\begin{abstract}
We study the mixing of the scalar glueball into the isosinglet mesons $f_0 (1370)$, $f_0(1500)$, and $f_0(1710)$ to describe the two-body decays to pseudoscalars. We use an effective Hamiltonian and employ the two-angle mixing scheme for $\eta$ and $\eta'$. In this framework, we analyse existing data and look forward to new data into $\eta$ and $\eta'$ channels. For now, the $f_0 (1710)$ has the largest glueball component and a sizable branching ratio into $\eta\eta'$, testable at BESIII.
\end{abstract}


\maketitle


\section{Introduction}

Glueballs are arguably the most important unconfirmed prediction of quantum chromodynamics (QCD) (see Refs.~\cite{Mathieu:2008me,Crede:2008vw,Ochs:2013gi} for reviews). Lattice QCD calculations predict the lowest-lying glueball around $1.6$--$\unit[1.7]{GeV}$, with quantum numbers $J^{PC} = 0^{++}$. It is expected to mix with $q\bar q$ states in the same mass region, resulting in more $0^{++}$ states than naively expected from $q\bar q$ spectroscopy. There seems to be some consensus that the states $f_0 (1370)$, $f_0 (1500)$, and $f_0 (1710)$ are the relevant mass eigenstates made up from $n\bar n \equiv (u\bar u + d \bar d )/\sqrt{2}$, $s\bar s$, and the $0^{++}$ glueball $G$ (see Tab.~\ref{tab:f0s} for masses and widths). 
We note however that the lightest state $f_0 (1370)$ is very poorly known, due to overlap with nearby states. 
To find the correct mixing structure one uses data on $f_0$ both from production (seeking ``glue-rich'' channels) and decay; for this it is important to identify processes that are particularly sensitive to the glueball couplings.

Gluons have a strong coupling to $\eta$ and $\eta'$ through the axial anomaly, which neatly describes the ratio $\Gamma (J/\psi\to \eta' \gamma)/\Gamma (J/\psi\to \eta \gamma)\sim 5$~\cite{Novikov:1979uy, Gershtein:1983kc,Ball:1995zv,Frere:1997xe}. For glueballs $G$ -- produced for example in $J/\psi\to G \gamma$ -- we expect $\Gamma (G\to \eta\eta) > \Gamma (G\to\pi\pi, KK)$ (expect the same from lattice QCD~\cite{Sexton:1995kd,Cheng:2015iaa}). See Tab.~\ref{tab:eta_ratios} for experimental decay rates of the potential glueball candidates $f_0 (1370,1500,1710)$ into $\eta\eta$ and $\eta\eta'$ from WA102 (soon to be updated with BESIII data).\footnote{See Refs.~\cite{Alde:1987mp,Alde:1988dz} for early experimental results on the $\eta$ channels and Ref.~\cite{Alde:1997ri} for $\pi$ channels.}
It is hard to see a coherent picture emerging in the data. In particular, all $f_0(1370)$ data are inconclusive/questionable to say the least. $f_0 (1500)$ is probably the best-studied state; only the $\eta\eta'$ channel is murky due to it being at the kinematic threshold. $f_0 (1710)$ data come mainly from WA102 and desperately require confirmation by BESIII. Our goal in this paper is to provide a framework to interpret upcoming data, which will hopefully clarify the picture.

The assignments of the three $f_0$ states to quark and glueball states are for the moment unclear, with different proposals in the literature~\cite{StrohmeierPresicek:1999yv}. Namely, Refs.~\cite{Amsler:1995tu,Amsler:1995td,Janowski:2011gt} suggested $f_0 (1500)$ to be dominantly glueball in nature, which does not fare that well anymore~\cite{Cheng:2015iaa}.
Refs.~\cite{Cheng:2015iaa,Cheng:2006hu} propose the mixing matrix
\begin{align}
\matrixx{f_0 (1370)\\ f_0 (1500)\\ f_0 (1710)} \equiv \matrixx{F_1\\ F_2 \\ F_3} = \matrixx{0.78 & 0.51 & -0.36\\ -0.54 & 0.84 & -0.03\\ 0.32 & 0.18 & 0.93} \matrixx{ n\bar n\\ s\bar s\\ G} ,
\end{align}
so $f_0 (1710)$ is mostly a glueball (same qualitative picture found e.g.~in Ref.~\cite{Janowski:2014ppa}).
The decay $f_0\to \eta\eta'$ is however not discussed, because it is close to threshold for the $f_0 (1500)$ -- and hence subject to systematic errors -- and inconsistently measured for the $f_0 (1700)$. Since the glueball is expected to have a strong connection to the $\eta$--$\eta'$ system, we will pay particular attention to these final states in this paper (see also Refs.~\cite{Narison:1988ts,Narison:1996fm}).

A strong argument for the glueball nature of $f_0(1710)$ comes from the $J/\psi$ decay rates~\cite{Gui:2012gx,Dobbs:2015dwa,Cheng:2015iaa}
\begin{align}
\frac{\Gamma (J/\psi \to \gamma f_0 (1710))}{\Gamma (J/\psi \to \gamma f_0 (1500))} = 10.5\pm 6.5
\end{align}
(where we used the $\pi\pi$ channel~\cite{Dobbs:2015dwa}),
which is expected to be enhanced for glueballs due to $J/\psi\to \gamma g g \to \gamma G$. One also finds~\cite{Dai:2015cwa}
\begin{align}
\frac{\Gamma (J/\psi \to \gamma f_0 (1370))}{\Gamma (J/\psi \to \gamma f_0 (1710))} = 0.51 \pm 0.41\,,
\end{align}
which fits well into that picture.

One more guideline for the identification of (predominantly) glueball states may be the total width, with a direct
decay to quarks expected to be somewhat suppressed (by the OZI rule or large-$N$ calculations). This argument is however
difficult to implement at the level of two-meson decays, as the total cross section may very well be dominated by more
complicated final states (in particular $f_0\to 4\pi$). 
Furthermore, some studies suggest rather broad glueballs~\cite{Ellis:1984jv}, so we will not impose theoretical conditions on the glueball width in this study.

\begin{table}
	\centering
	\renewcommand{\arraystretch}{1.8}
		\begin{tabular}{lll}
		\hline\hline
		State & Mass [MeV] & Width [MeV]\\		
		\hline		
			$a_0 (1450)$ & $1474\pm 19$ & $265\pm 13$\\
		\hline
			$f_0 (1370)$ & $1350\pm 150$ & $350\pm 150$\\
			$f_0 (1500)$ & $1505\pm 6$ & $109\pm 7$\\
			$f_0 (1710)$ & $1722\pm 6$ & $135\pm 7$\\
		\hline
		$\eta$  & $547.86\pm 0.02$ &  $\left(1.31\pm 0.05\right) \times 10^{-3}$\\
		$\eta'$ & $957.78\pm 0.06$ & $0.23\pm 0.02$\\
			\hline\hline
		\end{tabular}
		\caption{Some relevant masses and widths~\cite{Agashe:2014kda}.
		\label{tab:f0s}}
\end{table}

\begin{table}
	\centering
	\renewcommand{\arraystretch}{1.8}
		\begin{tabular}{llll}
		\hline\hline
		Decay ratio & Data & Fit 1 & Fit 2\\
		\hline
			$\frac{\Gamma ( a_0 (1450)\to KK)}{\Gamma ( a_0 (1450)\to \pi\eta)}$ & $0.88\pm 0.23$~\cite{Agashe:2014kda}* & $0.75^{+0.08}_{-0.06}$ & $0.81^{+0.08}_{-0.07}$ \\
			$\frac{\Gamma ( a_0 (1450)\to \pi\eta')}{\Gamma ( a_0 (1450)\to \pi\eta)}$ & $0.35\pm 0.16$~\cite{Agashe:2014kda} & $0.44$ & $0.44$\\
		\hline
			$\frac{\Gamma ( f_0 (1370)\to \pi\pi)}{\Gamma ( f_0 (1370)\to KK)}$ & $12.5\pm 12.5$~\cite{Ablikim:2004wn}* & $11^{+99}_{-8}$ & $10.1^{+41.9}_{-6.4}$\\
			$\frac{\Gamma ( f_0 (1500)\to \pi\pi)}{\Gamma ( f_0 (1500)\to KK)}$ & $4.07\pm 0.43$~\cite{Agashe:2014kda}* & $3.99^{+0.94}_{-0.65}$ & $4.04^{+1.48}_{-0.89}$\\
			$\frac{\Gamma ( f_0 (1710)\to \pi\pi)}{\Gamma ( f_0 (1710)\to KK)}$ & $0.41^{+0.11}_{-0.17}$~\cite{Agashe:2014kda}* & $0.46^{+0.04}_{-0.05}$ & $0.20^{+0.03}_{-0.03}$\\
		\hline
			$\frac{\Gamma ( f_0 (1370)\to \eta\eta)}{\Gamma ( f_0 (1370)\to \pi\pi)}$ & $0.03_{-0.03}^{+0.04}$ & $0.03^{+0.08}_{-0.03}$ & $0.04^{+0.08}_{-0.04}$\\
			$\frac{\Gamma ( f_0 (1500)\to \eta\eta)}{\Gamma ( f_0 (1500)\to \pi\pi)}$ & $0.145\pm 0.027$~\cite{Agashe:2014kda}* & $0.14^{+0.02}_{-0.02}$ & $0.14^{+0.03}_{-0.03}$\\
			$\frac{\Gamma ( f_0 (1710)\to \eta\eta)}{\Gamma ( f_0 (1710)\to \pi\pi)}$ & $1.17^{+0.61}_{-0.48}$ & $0.44^{+0.10}_{-0.08}$ & $0.96^{+0.32}_{-0.22}$\\
		\hline
			$\frac{\Gamma ( f_0 (1370)\to \eta\eta)}{\Gamma ( f_0 (1370)\to KK)}$ &  $0.35\pm 0.30$~\cite{Barberis:2000cd,Fariborz:2015era}* & $0.36^{+9.9}_{-0.36}$ & $0.39^{+4.4}_{-0.37}$\\
			$\frac{\Gamma ( f_0 (1500)\to \eta\eta)}{\Gamma ( f_0 (1500)\to KK)}$ &  $0.59\pm 0.12$~\cite{Agashe:2014kda} & $0.56^{+0.19}_{-0.14}$ & $0.58^{+0.26}_{-0.16}$\\
			$\frac{\Gamma ( f_0 (1710)\to \eta\eta)}{\Gamma ( f_0 (1710)\to KK)}$ &  $0.48\pm 0.15$~\cite{Barberis:2000cd,Agashe:2014kda}* & $0.20^{+0.03}_{-0.03}$ & $0.19^{+0.03}_{-0.03}$\\
		\hline
			$\frac{\Gamma ( f_0 (1500)\to \eta\eta')}{\Gamma ( f_0 (1500)\to \eta\eta)}$ & $0.38\pm 0.16$~\cite{Agashe:2014kda}* & $0.53^{+0.13}_{-0.11}$ & $0.47^{+0.14}_{-0.12}$\\
			$\frac{\Gamma ( f_0 (1710)\to \eta\eta')}{\Gamma ( f_0 (1710)\to \eta\eta)}$ &  $< 0.08$~\cite{Barberis:2000cd} & $0.64^{+0.22}_{-0.22}$ & $0.32^{+0.20}_{-0.18}$\\
			\hline
			$\Br ( f_0 (1500)\to \pi\pi)$ &  $0.349\pm 0.023$~\cite{Agashe:2014kda}* & $0.36^{+0.08}_{-0.07}$ & $0.35^{+0.07}_{-0.06}$\\
			$\Br ( f_0 (1710)\to KK)$ &  $0.36\pm 0.12$~\cite{Albaladejo:2008qa}* & $0.37^{+0.06}_{-0.06}$ & $0.44^{+0.07}_{-0.07}$\\
		\hline
		$\chi^2_\text{min}/\text{d.o.f.}$ & & $4.9/3$ & $6.4/3$ \\
		\hline
		\hline
		\end{tabular}
		\caption{Data without a reference have been obtained by combining referenced data. Data with an asterisk are fitted, everything else are predictions.
		\label{tab:eta_ratios}}
\end{table}

\section{Effective Hamiltonian}

\begin{figure*}[t]
\includegraphics[width=0.7\textwidth]{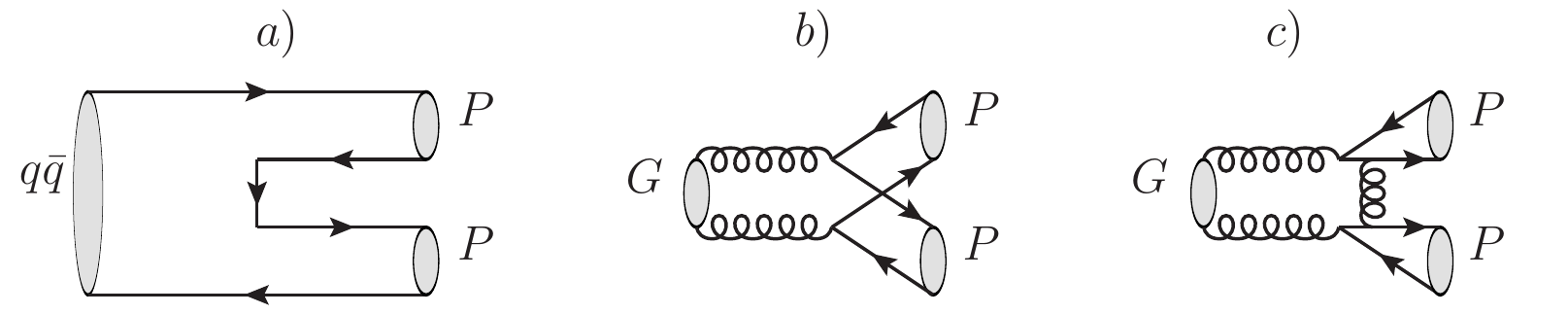}
\caption{Diagrammatic representation of the effective couplings $h_1$ (a), $h_2$ (b), and $h_3$ (c).}
\label{fig:glueball_decay}
\end{figure*}

To calculate the $f_0$ decays, we employ the effective Hamiltonian~\cite{Salomone:1980sp,Schechter:1982tv, Gao:1999br}
\begin{align}
\mathcal{H} &= h_1\, \tr \left[ X_F P P \right] + h_2\, G\,\tr \left[ P P \right] + h_3\, G\, \tr \left[P \right]\tr \left[P\right] ,
\label{eq:effective_hamiltonian}
\end{align}
with the pseudoscalar $SU(3)_f$ nonet
\begin{align}
P = \matrixx{\frac{\pi^0}{\sqrt{2}} + \frac{\eta_8}{\sqrt{6}} + \frac{\eta_0}{\sqrt{3}} & \pi^+ & K^+ \\ \pi^- & -\frac{\pi^0}{\sqrt{2}} + \frac{\eta_8}{\sqrt{6}} + \frac{\eta_0}{\sqrt{3}} & K^0 \\ K^- & \bar K^0 & - \frac{2 \eta_8}{\sqrt{6}} + \frac{\eta_0}{\sqrt{3}}} ,
\end{align}
and the scalar $q\bar q$ states
\begin{align}
X_F = \matrixx{ u\bar u & & \\ & d\bar d & \\ & & s\bar s} .
\end{align}
See Fig.~\ref{fig:glueball_decay} for a diagrammatic representation of these operators (see also Ref.~\cite{StrohmeierPresicek:1999yv}). (Note that Ref.~\cite{Giacosa:2005zt} proposes an entirely different effective Hamiltonian of chiral perturbation theory; for the time being, we prefer not to rely on chiral perturbation in the energy range considered -- more on this later about chiral suppression.)

Assuming isospin invariance, we will adopt the standard mixing scheme
\begin{align}
\matrixx{a_0 (1450)\\f_0(1370)\\f_0(1500)\\f_0(1710)} = \matrixx{ 1 & \\ & U}\matrixx{ \tfrac{1}{\sqrt{2}} & -\tfrac{1}{\sqrt{2}} & & \\ \tfrac{1}{\sqrt{2}}& \tfrac{1}{\sqrt{2}}  & & \\ & & 1 & \\ & & & 1} \matrixx{u\bar u\\ d\bar d\\ s\bar s\\ G} ,
\label{eq:mixing_matrix}
\end{align}
where $U$ is a real orthogonal $3\times 3$ matrix describing the mixing of $n\bar n$, $s\bar s$, and $G$ into the $f_0$ states, further abbreviated by $(F_1,F_2,F_3)\equiv (f_0(1370),f_0(1500),f_0(1710))$.

With the Hamiltonian from Eq.~\eqref{eq:effective_hamiltonian} one can easily calculate all the $f_0\to \pi\pi$, $KK$, $\eta\eta$ decays~\cite{Cheng:2006hu}, as well as the $f_0\to\eta \eta'$ decays of interest to us (which has been done already in Ref.~\cite{Li:2000yn}, but only using $h_{1,2}$).
The question is now how to include $SU(3)_f$-breaking effects in Eq.~\eqref{eq:effective_hamiltonian} -- which should exist due to $m_s\gg m_{u,d}$. Some remarks are in order:

\begin{itemize}

	\item The coupling $h_1$ (Fig.~\ref{fig:glueball_decay} a) describes the OZI-allowed decay that should dominate the $q\bar q$ decays. The creation of an $s\bar s$ from the vacuum is modified relative to $u\bar u$ by a factor $r_a \sim 1$~\cite{Cheng:2006hu}, which we include explicitly in the amplitudes of $h_1$.
	
	\item $h_2$ corresponds to the dominant glueball couplings to the nonet (Fig.~\ref{fig:glueball_decay} b). A flavour-democratic coupling of glueballs to quarks would lead to the ratios
\begin{align*}
\qquad &\M (G \to \pi\pi) : \M (G\to K K ) : \M (G \to \eta\eta)\\  
&\  :\M (G\to \eta'\eta') :\M (G\to \eta\eta') = 1:1:1:1:0\,,
\end{align*}
and hence decay rates $3:4:1:1:0$ modulo phase space~\cite{Ochs:2013gi}.
Some of the relevant diagrams, however, contain the subprocess $gg\to q\bar q$ (differently attached gluons in Fig.~\ref{fig:glueball_decay} b) and it has been argued that this implies a chiral suppression $\sigma (gg \to q\bar q)\propto m_q$~\cite{Carlson:1980kh,Chanowitz:2005du}. How this suppression propagates to the hadronized $G\to PP$ rate is unfortunately unknown. A proportionality of $\M (G\to PP)$ to current-quark masses is highly improbable for several reasons: (i) even accepting a Feynman-diagram picture of these strong interactions, the gluons may land on different quark lines,\footnote{Meaning $\sigma (gg \to q\bar q)\propto m_q$ is in any case only one possible subprocess.} (ii) even accepting the chiral suppression, we know that chiral symmetry is broken by confinement itself, leading to constituent quark masses, (iii) alternate calculations based on perturbative QCD (equally out of their domain of application as chiral perturbation) suggest instead the decay constants $f_{\pi}$, $f_K$, etc., as suppression~\cite{Chao:2005si,Cheng:2006hu}, and (iv) completely different arguments would use the overlap of wave functions at the origin as a factor governing the flavour nonuniversality.\footnote{In Ref.~\cite{Brunner:2015yha}, a coupling of the form $M_P^2 P^2 G$ is motivated within the Witten--Sakai--Sugimoto model for the singlet pseudoscalar $\eta_0$ and subsequently extended to all $P$.}

(Note also that the chiral suppression argument of the coupling of quarks to two gluons does not apply to the case of pseudoscalars through the quantum anomaly.) We thus leave the extent of the suppression (if any) free for now, and introduce different couplings for the $PP$ states to take the chiral dependence into account, following again Ref.~\cite{Cheng:2006hu}:
\begin{align}
\begin{split}
h_2\, G\,\tr \left[ P P \right] &\to  G\left[ h_{2,\pi} \left((\pi^0)^2 + 2 \pi^+ \pi^-\right) \right.\\
&\qquad  + 2 h_{2,K} \left(K^0\bar K^0 + K^+ K^-\right) \\
&\qquad \left. + h_{2,0} (\eta_0)^2 + h_{2,8} (\eta_8)^2\right] .
\end{split}
\end{align}

\item $h_3$ (Fig.~\ref{fig:glueball_decay} c) describes the $G\eta_0\eta_0$ coupling that can be associated with the anomaly~\cite{Salomone:1980sp,Schechter:1982tv} and is of particular interest to our study of $\eta\eta'$ final states. (In Ref.~\cite{Li:2000cj} this is taken one step further and promoted to a glueball--glueball--glueball coupling followed by a mixing of $\eta$ and $\eta'$ with the $0^{-+}$ glueball~\cite{Cheng:2008ss}.)
The full coupling is $\mathcal{H} \supset (h_{2,0} + 3 h_3) G (\eta_0)^2$, but we will fix $h_{2,0}$ instead of fitting it, so $h_3$ is indeed a free parameter for us.

\item Couplings $\tr \left[ X_F \right] \tr \left[P P \right]$, $\tr \left[ X_F P\right] \tr \left[P \right] $, and $\tr \left[ X_F\right] \tr \left[P\right] \tr \left[P \right]$ would correspond to OZI-suppressed diagrams (see Ref.~\cite{He:2006tw}) and are neglected here as usual.

\item The free mixing matrix $U$ of Eq.~\eqref{eq:mixing_matrix} describes in itself $SU(3)_f$ violation coming from the nondiagonal mass matrix of the $q\bar q $ and $G$ states. We will not assume a structure in this mass matrix (see e.g.~Refs.~\cite{Weingarten:1996pp,Lee:1999kv}) but rather determine $U$ directly from the $f_0$ decays, keeping the mass eigenvalues and widths fixed to their PDG values (Tab.~\ref{tab:f0s}).

\end{itemize}

\subsection{\texorpdfstring{$\eta$}{eta}--\texorpdfstring{$\eta'$}{eta-prime} mixing}

Most studies of $f_0$ decays assume the $\eta$--$\eta'$ system (Tab.~\ref{tab:f0s}) to be an orthogonal transformation of the $SU(3)_f$ states,
\begin{align}
\matrixx{\eta\\ \eta'}= \matrixx{\cos \theta_P & -\sin \theta_P\\\sin\theta_P & \cos\theta_P} \matrixx{\eta_8\\ \eta_0} ,
\end{align}
with flavour eigenstates $\eta_0 = (u\bar u + d\bar d + s\bar s)/\sqrt{3}$, $\eta_8 = (u\bar u + d\bar d -2 s\bar s)/\sqrt{6}$, and mixing angle $\theta_P = -11.4^\circ$~\cite{Agashe:2014kda} (see also Ref.~\cite{Akhoury:1987ed}).

It has been shown that this approach is a little too na\"ive, because the large hierarchy $m_\eta\ll m_{\eta'}$, together with renormalization effects, complicate matters~\cite{Leutwyler:1997yr,Escribano:1999nh,Mathieu:2010ss}. In effect, one should rather use a general (invertible) transformation matrix, which has four real parameters,
\begin{align}
\matrixx{\eta\\ \eta'}&= \frac{1}{F}\matrixx{F_8\cos \theta_8 & -F_0\sin \theta_0\\F_8 \sin\theta_8 & F_0\cos\theta_0} \matrixx{\eta_8\\ \eta_0} ,
\end{align}
and is called the two-mixing-angle scheme.\footnote{Using the quark basis $\eta_{q,s}$ instead of the flavor basis $\eta_{8,0}$ in the two-angle scheme leads to two mixing angles $\theta_{q,s}$ that are accidentally close to each other, allowing (at current precision) for a one-angle description $\theta_q = \theta_s$. Since this is obviously a basis-dependent effect, we will keep the two-angle formalism here.} 
(It reduces to the one-angle scheme in the limit $F_0 = F_8 = F$, $\theta_0 = \theta_8 = \theta_P$.)
Here, $F= \unit[92.2]{MeV}$ is a normalization constant, and a fit gives~\cite{Feldmann:1998vh}
\begin{align}
F_8/F &= 1.26\pm 0.04\,, &
\theta_8 &= (-21.2\pm 1.6)^\circ ,\\
F_0/F &= 1.17\pm 0.03 \,, &
\theta_0 &= (-9.2\pm 1.7)^\circ .
\end{align}
(Other methods of determining these parameters give similar values; see e.g.~Refs~\cite{Feldmann:1999uf,Escribano:2005qq,Chen:2012vw,Escribano:2015nra}.)

Since this two-angle scheme is rather successful and stable in describing $\eta$--$\eta'$ data, we will keep the above parameters fixed to the best-fit values.
For $a_0 (1450)$, we then immediately arrive at the ratios
\begin{align}
\begin{split}
\frac{\Gamma ( a_0 (1450)\to \pi\eta')}{\Gamma ( a_0 (1450)\to \pi\eta)} &= \left(\frac{2 F_8 \cos \theta_8 + \sqrt{2} F_0 \sin\theta_0}{2 F_8 \sin \theta_8 - \sqrt{2} F_0 \cos\theta_0}\right)^2 \frac{p_{\eta'}}{p_\eta}\\
&\simeq 0.44\,,
\end{split}
\end{align}
and
\begin{align}
\begin{split}
\frac{\Gamma ( a_0 (1450)\to KK)}{\Gamma ( a_0 (1450)\to \pi\eta)} &= \left(\frac{\sqrt3 r_a F_0 F_8 \cos (\theta_0 - \theta_8)/F}{\sqrt{2} F_0 \cos\theta_0-2F_8\sin \theta_8}\right)^2 \frac{p_{K}}{p_\eta}\\
&\simeq 0.84 \,r_a^2\,,
\end{split}
\end{align}
perfectly compatible with experimental results for $r_a\simeq 1$ (Tab.~\ref{tab:eta_ratios}).

The $f_0$ decays are more involved due to the mixing matrix $U$, which we will determine by a numerical fit in the next section. The formulae for the decay rates are collected in the Appendix.

\subsection{Fit}

Having defined our Hamiltonian, we can perform our own fit to the decay data in Tab.~\ref{tab:eta_ratios}, using a simple $\chi^2$ function. Note that we will keep the masses fixed (Tab.~\ref{tab:f0s}) and do not impose a structure on the $f_0$ mass matrix in flavour basis (given by $U^T \diag (M_{F_1},M_{F_2},M_{F_3}) U$). Since there is unfortunately not enough reliable data to obtain a statistically significant result, our best-fit values have to be taken with a grain of salt.

In order to calculate the rate $f_0 (1500)\to \eta\eta'$, which is at threshold because $M_{\eta} + M_{\eta'}= \unit[1505.6]{MeV} \gtrsim M_{F_2}$ (see Tab.~\ref{tab:f0s}), we integrate the partial widths with a Breit--Wigner distribution:
\begin{align}
\begin{split}
&\frac{\Gamma (F_2 \to \eta\eta')}{\Gamma (F_2\to \eta\eta)} = \frac{|\mathcal{M}(F_2 \to \eta\eta')|^2}{|\mathcal{M}(F_2 \to \eta\eta)|^2}\\
&\times \frac{\int_{M_{\eta} + M_{\eta'}}^{M_{F_2} + \Gamma_{F_2}}\dd E \, p(E,M_{\eta},M_{\eta'})/[(E^2-M_{F_2}^2)^2 + M_{F_2}^2 \Gamma_{F_2}^2]}{\int_{M_{F_2} - \Gamma_{F_2}}^{M_{F_2} + \Gamma_{F_2}}\dd E \, p(E,M_{\eta},M_{\eta})/[(E^2-M_{F_2}^2)^2 + M_{F_2}^2 \Gamma_{F_2}^2]}.
\end{split}
\end{align}
Here,
\begin{align}
p(E,M_1,M_2) \equiv \frac{\sqrt{[E^2-(M_1 + M_2)^2][E^2-(M_1-M_2)^2]}}{2 E}
\end{align}
is the momentum of a daughter particle in a two-body decay, where the initial particle has ``mass'' $E$.
Since the $F_2\to \eta\eta'$ decay relies on a slightly off-shell $F_2$, we integrated the $\eta\eta'$ and $\eta\eta$ decay rates with the highly peaked Breit--Wigner distribution over $[M_{F_2} - \Gamma_{F_2},M_{F_2} + \Gamma_{F_2}]$, which reduces to $[M_{\eta} + M_{\eta'},M_{F_2} + \Gamma_{F_2}]$ for $\eta\eta'$ due to kinematics.
Numerically, one finds
\begin{align}
\begin{split}
\frac{\Gamma (F_2 \to \eta\eta')}{\Gamma (F_2 \to \eta\eta)} = 0.14\times \frac{|\mathcal{M}(F_2 \to \eta\eta')|^2}{|\mathcal{M}(F_2 \to \eta\eta)|^2} ,
\end{split}
\end{align}
fairly stable against small changes in the domain of integration.

For the decay ratios of $J/\psi$ we make the simple assumption that the decay is dominated by $J/\psi\to gg\gamma \to G\gamma$, leading to the simple formula
\begin{align}
\frac{\Gamma (J/\psi \to \gamma F_i)}{\Gamma (J/\psi \to \gamma F_j)} = \frac{|U_{i3}|^2}{|U_{j3}|^2} \frac{p_{F_i}}{p_{F_j}}\,,
\end{align}
but we will not include these ratios in our fits. See e.g.~Refs.~\cite{Close:2005vf,Chatzis:2011qz} for dedicated studies.

\subsubsection{Fit 1: Flavour blind}

Taking for simplicity a flavour-blind $G$ coupling $h_2 \equiv h_{2,\pi} = h_{2,K} = h_{2,0} = h_{2,8}$ and the two-angle $\eta$--$\eta'$ values from above we can fit $h_{1,2,3}$ and the mixing matrix $U$, parametrized via three angles $\theta_{ij}$ (see Eq.~\eqref{eq:U}). We find\footnote{The $1\sigma$ range for parameter $x$ has been obtained by fixing all other parameters to their best-fit values and then solving $\chi^2 (x) = \chi^2_\mathrm{min} +1$ for $x$.}
\begin{align}
r_a &= 0.94^{+0.05}_{-0.04}\,, & h_1 &= \unit[(698\pm 30)]{MeV}\,,\\
h_2/h_1 &= 1.12^{+0.03}_{-0.04}\,, & h_3/h_1 &= 1.5^{+0.2}_{-0.3}\,,
\end{align}
for the coupling constants and
\begin{align}
\theta_{23} = 0.67\pm 0.03\,, &&
\theta_{12} = 3.56\pm 0.04\,, &&
\theta_{13} = 0.38^{+0.01}_{-0.02}\,, 
\end{align}
for the mixing angles, resulting in the mixing structure
\begin{align}
\matrixx{F_1\\F_2\\F_3} &= 
\left(
\begin{array}{ccc}
 -0.85 & -0.38 & 0.37 \\
 0.53 & -0.63 & 0.57 \\
 0.02 & 0.68 & 0.73 \\
\end{array}
\right)
\matrixx{n\bar n\\ s\bar s\\G}\\
&=
\left(
\begin{array}{ccc}
 -0.91 & -0.18 & 0.37 \\
 0.07 & 0.82 & 0.57 \\
 0.41 & -0.55 & 0.73 \\
\end{array}
\right) \matrixx{\text{singlet}\\ \text{octet} \\G} ,
\label{eq:singlet-octet-G}
\end{align}
meaning that the largest glueball component is in $f_0(1710)$.
The best-fit decay widths and ratios are given in Tab.~\ref{tab:eta_ratios}.

We also gave the mixing matrix for the singlet--octet--$G$ basis in Eq.~\eqref{eq:singlet-octet-G}, which seems to be a better starting point than $n\bar n$--$s\bar s$--$G$, since the mixing angles in this basis are smaller. $f_0(1500)$ is dominantly a flavour octet, similar to Refs.~\cite{Cheng:2006hu,Cheng:2015iaa}.
While we have made no considerations on the mass mixing matrix, and have determined the mixing
matrix $U$ from decay properties, it is an interesting consistency check to verify to which original
mass matrix $M =U^T M_\mathrm{diag} U$ in the flavour basis $n\bar n$, $s\bar s$, $G$ our fit corresponds:
\begin{align}
M &= \left(
\begin{array}{ccc}
 1393.09 & -46.93 & 51.41 \\
 -46.93 & 1584.15 & 129.72 \\
 51.41 & 129.72 & 1599.76 \\
\end{array}
\right) \unit{MeV},
\end{align}
or 
\begin{align}
\left(
\begin{array}{ccc}
 1412.53 & -74.43 & 116.87 \\
 -74.43 & 1564.71 & -76.24 \\
 116.87 & -76.24 & 1599.76 \\
\end{array}
\right)\unit{MeV},
\end{align}
in the singlet--octet--$G$ basis. It is reassuring to see that the $SU(3)$-breaking elements are indeed smaller than the conserving (diagonal) ones and that no fine-tuned cancellations occur. However, the ``octet mass'' $\unit[1565]{MeV}$ comes out surprisingly large compared to the states $K^* (1430)$ and $a_0 (1450)$.

The ratios for $J/\psi$ decays into $\gamma F_j$ show the qualitative behavior we expected from our initial considerations,
\begin{align*}
\frac{\Gamma (J/\psi \to \gamma f_0 (1370))}{\Gamma (J/\psi \to \gamma f_0 (1710))} &= 0.3\,,\\
\frac{\Gamma (J/\psi \to \gamma f_0 (1710))}{\Gamma (J/\psi \to \gamma f_0 (1500))} &= 1.5\,.
\end{align*}
We find, however, a very small $f_0 (1370)$ width into $PP$,
\begin{align}
\Gamma (f_0 (1370)\to PP) &\simeq \unit[2]{MeV} \,,
\end{align}
dominated by the $\pi\pi$ final state.
From Eq.~\eqref{eq:Fitopipi} we see that this small rate is due to an interference of the $n\bar n$ and $G$ amplitudes, reducing the rate by an order of magnitude.
While we expect this rate to be subdominant to $f_0 (1370)\to 4\pi$~\cite{Agashe:2014kda}, and the full width is known only very imprecisely, this might still be the most worrying result of our fit. (In Ref.~\cite{Bugg:2007ja} it is argued that the $4\pi$ channel is subdominant to the $2\pi$ channel.)

\subsubsection{Fit 2: Chiral suppression}

Since we expect a chiral suppression in the $G\to PP$ decays, we perform a second fit with the ansatz
\begin{align}
h_{2,\pi} &= h_2\,, &  h_{2,K} &= (f_K/f_\pi)^2 h_2\,, \\
h_{2,0} &= (F_0/F)^2 h_2\,, & h_{2,8} &= (F_8/F)^2 h_2\,,
\end{align}
(motivated partly by Refs.~\cite{Chao:2005si,Cheng:2006hu}), which gives
\begin{align}
r_a &=0.98^{+0.05}_{-0.04}\,, & h_1 &= \unit[(910\pm 40)]{MeV}\,,\\
h_2/h_1 &=-0.59\pm 0.02\,, & h_3/h_1 &= -1.0\pm 0.2\,,
\end{align}
for the coupling constants and
\begin{align}
\theta_{23} = 0.60\pm 0.02\,, &&
\theta_{12} = 0.57\pm 0.04\,, &&
\theta_{13} = 0.51^{+0.01}_{-0.02}\,, 
\end{align}
for the mixing angles, resulting in the mixing structure
\begin{align}
\matrixx{F_1\\F_2\\F_3} &= 
\left(
\begin{array}{ccc}
 0.74 & 0.47 & 0.49 \\
 -0.68 & 0.55 & 0.49 \\
 -0.04 & -0.69 & 0.72 \\
\end{array}
\right)
\matrixx{n\bar n\\ s\bar s\\G}\\
&=
\left(
\begin{array}{ccc}
 0.87 & 0.04 & 0.49 \\
 -0.24 & -0.84 & 0.49 \\
 -0.43 & 0.54 & 0.72 \\
\end{array}
\right) \matrixx{\text{singlet}\\ \text{octet} \\G} ,
\end{align}
with only minor quantitative differences to the flavour-blind fit.
The mass matrix in the $n\bar n$--$s\bar s$--$G$ basis is given by
\begin{align}
\left(
\begin{array}{ccc}
 1421.53 & -48.29 & -61.19 \\
 -48.29 & 1574.06 & -143.85 \\
 -61.19 & -143.85 & 1581.41 \\
\end{array}
\right) \unit{MeV},
\end{align}
and 
\begin{align}
\left(
\begin{array}{ccc}
 1426.85 & -55.81 & -133.02 \\
 -55.81 & 1568.74 & 82.13 \\
 -133.02 & 82.13 & 1581.41 \\
\end{array}
\right)\unit{MeV},
\end{align}
in the singlet--octet--$G$ basis. 
In addition, we obtain the partial width
\begin{align}
\Gamma (f_0 (1370)\to PP) &= \unit[4]{MeV} \,,
\end{align}
and the ratios
\begin{align*}
\frac{\Gamma (J/\psi \to \gamma f_0 (1370))}{\Gamma (J/\psi \to \gamma f_0 (1710))} &= 0.5\,,\\
\frac{\Gamma (J/\psi \to \gamma f_0 (1710))}{\Gamma (J/\psi \to \gamma f_0 (1500))} &= 2\,.
\end{align*}
The same comments as above apply regarding the width of $f_0(1370)$.


\section{Discussion and Conclusion}

With the present state of data, the above fits can obviously not convincingly decide on the main issues of glueball spectroscopy, namely the composition of the $f_0$ states and their decay properties. We consider the present attempt mainly as a preparation for the interpretation of future data.
Nevertheless, in both fits we find that $f_0 (1710)$ has the largest admixture of the $0^{++}$ glueball, whereas $f_0 (1500)$ comes out close to the octet flavour structure.
The very small rate of $f_0 (1370)\to PP$ is particularly concerning, even though $f_0 (1370)\to 4 \pi$ is expected to make up most of the width. 
The coupling of the glueball to $\eta_0$ plays an important role in both fits and is numerically large, making $\eta$ and $\eta'$ final states crucial testing grounds.
We certainly expect $f_0 (1710)\to \eta\eta'$ to be visible in BESIII, which will hopefully clarify some of the issues. It may also be surprising to some that we obtain a better fit with flavour-blind glueball couplings.

We stress that the use of the two-angle scheme for $\eta$--$\eta'$ mixing is important when discussing the $f_0\to \eta\eta(')$ decay modes, as the usually employed one-angle scheme is inconsistent.

\acknowledgments{We thank Simon Mollet for discussions and Hai-Yang Cheng for additional information regarding Ref.~\cite{Cheng:2006hu}. JMF thanks the GAMS group for introducing him to this subject and for many discussions in the past.  This work is funded in part by IISN and by Belgian Science Policy (IAP VII/37).}


\appendix

\section{Decay rates and ratios}
\label{sec:decay_rates}

The $3\times 3$ orthogonal mixing matrix $U$ that describes the mixing of the $f_0$ states and the $q\bar q$ and glueball states is defined by
\begin{align}
\matrixx{f_0 (1370)\\ f_0 (1500)\\ f_0 (1710)} \equiv \matrixx{F_1\\ F_2 \\ F_3} = U\matrixx{ n\bar n\\ s\bar s\\ G} ,
\end{align}
and for the numerical fit parametrized by three mixing angles $\theta_{12}$, $\theta_{23}$, and $\theta_{13}$ using the common PDG notation
\begin{align}
U = \matrixx{c_{12} c_{13} & s_{12} c_{13} & s_{13}\\
	-c_{23} s_{12}- s_{23} s_{13} c_{12} & c_{23} c_{12}- s_{23} s_{13} s_{12} & s_{23} c_{13}\\
	s_{23}s_{12}- c_{23} s_{13} c_{12} & -s_{23} c_{12}- c_{23} s_{13} s_{12} & c_{23} c_{13}}
\label{eq:U}
\end{align}
with $c_{ij}\equiv \cos \theta_{ij}$ and $s_{ij}\equiv \sin \theta_{ij}$.
For the decay rates, we define $\rho_{2,X}\equiv h_{2,X}/h_1$ and $\rho_3 \equiv h_3/h_1$, leading to
\begin{align}
\Gamma (F_i \to \pi\pi) &= \frac{1}{8\pi} \frac{3 h_1^2}{2}\left(\sqrt{2} U_{i1} + 2 \rho_{2,\pi} U_{i3}\right)^2 \frac{p_\pi}{m_i^2} \,.
\label{eq:Fitopipi}
\end{align}
Here and below, $p_X = |\vec{p}_X|$, where $\vec{p}_X$ denotes the three-momentum of $X$ in the two-body decay $F_i\to XY$.
Introducing
\begin{align}
A_i &\equiv \sqrt{2} U_{i1} + 4 r_a U_{i2} + 6 \rho_{2,8} U_{i3}\,,\\
B_i &\equiv U_{i1} - \sqrt{2} r_a U_{i2}\,,\\
C_i &\equiv 2\left[ \sqrt{2} U_{i1} + r_a U_{i2} + 3 (\rho_{2,0} + 3 \rho_3) U_{i3}\right] ,
\end{align}
we find the ratios
\begin{widetext}
\begin{align}
\frac{\Gamma (F_i\to \pi\pi)}{\Gamma (F_i \to KK)} &= \frac{3}{2}\left(\frac{\sqrt{2} U_{i1} + 2 \rho_{2,\pi} U_{i3}}{r_a U_{i1}+\sqrt{2} U_{i2} + 2\sqrt{2} \rho_{2,K} U_{i3}} \right)^2 \frac{p_\pi}{p_K} \,,\\
\frac{\Gamma (F_i\to \eta\eta)}{\Gamma (F_i \to \pi\pi)} &= \frac{1}{27}\left(\frac{F^2 \left[F_0^2 A_i \cos^2\theta_0 - 4 F_0 F_8 B_i \cos\theta_0 \sin\theta_8 + F_8^2 C_i \sin^2\theta_8\right]}{F_0^2 F_8^2 \cos^2 (\theta_0 - \theta_8)  \left[\sqrt{2} U_{i 1} + 2 \rho_{2,\pi} U_{i3}\right]}\right)^2\frac{p_{\eta}}{p_{\pi}} \,,\\
\frac{\Gamma (F_i\to \eta\eta')}{\Gamma (F_i \to \eta\eta)} &= \frac{1}{2}\left(\frac{F_0^2 A_i \sin 2\theta_0 +4 F_0 F_8 B_i \cos (\theta_0+\theta_8) - F_8^2 C_i \sin 2\theta_8}{F_0^2 A_i \cos^2\theta_0 - 4 F_0 F_8 B_i \cos\theta_0 \sin\theta_8 + F_8^2 C_i \sin^2\theta_8}\right)^2\frac{p_{\eta'}}{p_{\eta}} \, .
\end{align}
\end{widetext}

\bibliographystyle{utcaps_mod}
\bibliography{BIB}

\end{document}